# Construal Level and Cognitive Reflection in Newsvendor Games: Unveiling the Influence of Individual Heterogeneity on Decision-Making

## Kuldeep Singh [a], Sumanth Cheemalapati [b], George Kurian [c] and Prathamesh Muzumdar [d*]

*[a] Arkansas Tech University, United States.*
*[b] Dakota State University, United States.*
*[c] Eastern New Mexico University, Mexico.*
*[d] University of Texas at Arlington, United States.*

*Review Article*

## ABSTRACT

During the last decade, scholars have studied the behavior of decision-making in newsvendor settings and have identified numerous behavior patterns for deviating from normative behavior. However, there is a dearth of research which have examined the influence of individual heterogeneity on decision-making in newsvendor settings. This study examines the level of construal (Abstract and concrete) using construal level theory (CLT) on performance in newsvendor games. In addition, this study measures the cognitive reflection of individuals using cognitive reflection test (CRT) ex-ante to analyze the true impact of how people construe a problem and its impact on their decision-making.

*\*Corresponding author: Email: prathameshmuzumdar85@gmail.com;*

## 1. INTRODUCTION

In Supply chain management, the managers make decisions in an uncertain environment. One of the decisions is related to inventory management and Newsvendor is a classic example of how managers have to decide on order quantity with uncertain demand. In the newsvendor's problem, the demand is stochastic, and only stochastic distribution is known to the manager. The manager has to decide on stocking quantity before the selling period. If managers order much more than the demand, they won't be able to sell all the units and they have to dispose of the remaining units at salvage value (Kurian et al., 2024). On the other hand, if they order fewer units, then there will be stock out and lost sales.

The normative solution to the newsvendor problem is that decision-makers will order the optimal quantity. To test this phenomenon, several studies were done and it was observed that people deviate from the optimal solution (Muzumdar et al., 2024). In seminal work found that subject systematically deviates from optimal quantities that would maximize their expected profits both in low and high-margin settings. They called this phenomenon the "pull-to-center effect" and suggested that subjects' decisions are guided by heuristics such as anchoring and adjustment and inventory error minimization (Ogujiuba & Maponya, 2024). Since then, there have been several studies that highlighted different explanations for deviation from optimal solution. In addition, some studies have looked into the impact of other factors on decisions such as experience and feedback and group decision-making.

These studies observed similar results for order quantity but most of these studies reported average results and only a few studies have looked at variation in order quantity due to individual heterogeneity (Basyal et al., 2020). They examined the impact of culture on ordering decisions and the thought process of individuals by using a "think aloud "approach in a newsvendor setting. Moreover, research in another field such as cognitive psychology has hinted that individual differences may imply a cause of deviation between the normative and descriptive model of performance.

In supply chain management, researchers have also called for new theories to explain the individual biases of supply chain managers (Muzumdar, 2014f; Reavis & Stein, 2023). They studied order behavior from a bounded rationality perspective using the quantal model and called for additional research to study individual heterogeneity in decision-making and other researchers have made similar calls to examine other factors which may explain variation in decision making.

The main aim of this paper is to explain some of the individual heterogeneity observed in the newsvendor problem through the lens of construal level theory (CLT) (Muzumdar, 2014b). The construal level theory (CLT) from the psychology literature is used to show how the level of construal (abstract and concrete) is related to order variation and expected profits in a newsvendor setting (Reavis & Singh, 2023). According to construal level theory, the individual's choice of decision is based on whether people use the high-level construal (abstract mindset) or low-level construal (concrete mindset) at the time of decision making (Muzumdar, 2014d). We hypothesize that the level of construal (abstract and concrete) of individuals is related to outcome in a newsvendor setting. The experiment will be conducted to test the hypotheses with two settings (Singh et al., 2018). First, the focus will be on high margin setting to examine the variation in ordering behavior and expected profits due to different cognitive mechanisms (abstract and concrete) used by individuals. Second, we will repeat the experiment but with a low margin setting to see if the results are consistent as suggested by construal level theory. In addition, the Cognitive reflection test (CRT) will be measured ex-ante of all the subjects to control for the influence of Cognitive reflection. Moreover, this study will examine the type of behavior exhibited by subjects based on their construal level.

The contribution of this study is twofold. First, this is the first study that has used construal-level theory to explain the sum of heterogeneity in decision-making in news vendor settings. Second, this study highlights the importance of cognitive mechanisms and how they can influence the decision-making of individuals. Specifically, the human resource manager can devise a test to measure the construal level and these results can be used during the hiring or promotion of employees to key positions, which are critical for the success of organizations such as supply chain managers.

The remainder of this paper proceeds as follows. In section 2, the theoretical foundation of construal level theory (CLT), the newsvendor problem is discussed. In section 3, hypotheses are developed. Experimental design and methodology are explained in section 4. Section 5 discusses the theoretical and managerial implications and conclusion.

## 2. THEORETICAL DEVELOPMENT

### 2.1 Construal Level Theory

Construal theory posits that people view and approach the problem differently based on whether people have used high-level mental construal (abstract) or low-level construal (concrete) thinking process (Limaye et al., 2023). The level of abstraction (high or low) will alter the meaning of the event/ problem which will influence the outcome of the even. Construal-level theory sheds light on the how and why aspects of the decision-making process of people (Basyal et al., 2021a). At a high level of construal, the information about objects and event is conceptualized in abstract form (Muzumdar, 2015b). Abstract thinkers adopt a generalization view of the situation that allows for more alternatives and flexibility to act, which allows people to take control of the situation. High-level construal, "captures the superordinate central feature of an object or event," and individuals choose relevant information and omit the trivial information from the event [38]. For example, "waving of hand "will be construed as "showing friendliness" by the abstract thinker and he will discard the information about the details of one's hand.

As trivial information is omitted during abstract representation, they are simple, more unambiguous, and more prototypical (Elkassabgi et al., 2022). Abstract thinkers look beyond the current time and broaden their mental landscape by looking deep into the future and considering several and unique possibilities (Limaye et al., 2021). This implies that people with abstract thinking will explore various options before making a decision. The significance of judgment of more abstract construal is prominent because abstract thinkers use stereotypes information, generalized scripts, trait concepts, and causal explanations at the time of analyzing a situation (Basyal et al., 2021b). Abstract thinking enhances creativity, innovation, and insight to solve the problem (Muzumdar, 2012a). They conducted an experimental study to examine the effect of mental construal on creativity and found that individuals with abstract thinking performed better on tasks that involved the insight aspect of creativity.

Action identification theory argues that actions can be represented in several ways including superordinate and subordinate goals (Muzumdar & Kurian, 2019b). Actions that encompass superordinate goals are related to high-level construal and signify the importance of the why aspect of action (Choi & Park, 2022). Individuals with high levels of construal associate their actions with the big picture, motive, and primary goal of their action.

On the other hand, concrete thinkers (low-level construal) use a low level of abstraction to solve a problem. Concrete thinkers use, "individuating information, non-schematic details, and incoherent occurrences during the evaluation of problem" (Muzumdar, 2014e). During interpretation of event concrete thinkers focus more on secondary and incidental features of the event. Concrete thinkers consider contextualized, subordinate, incoherent qualities of an event (Singh et al., 2024). It is seen that individuals having a concrete perspective will identify an event (Olympic games) based on features such as flashy advertisements and presence of the celebrities (Muzumdar et al., 2024).

People with low-level construal focus on the feasibility aspect of choice when faced with choice alternatives which implies that they focus more on means to reach the end state. In their study, they found that the feasibility perspective was increased among individuals having low-level construal at the time of deciding a choice (Muzumdar, 2012b). Individuals with low level construal focus more on peripheral and goal-irrelevant concerns at the time of decision and people give more weight to peripheral features during judgment and decisions (Muzumdar et al., 2023). Concrete thinkers are directed toward the action process and focus on how aspect of action. Table 1 summarizes the main features of high and low levels of construal levels.

**Table 1. Summary of attribute of construal level**

| High-Level Construal | Lowe level Construal |
|---|---|
| Focus on the central feature | Focus on peripheral feature |
| Creative | Focus on goal irrelevant concern |
| Flexible | Action process oriented |
| Superordinate goal | Subordinate goal |
| Why aspect of action | How aspect of action |

## 2.2 Newsvendor Problem

In a newsvendor setting, the seller has to decide how much to order (q) to satisfy the stochastic demand D during a single selling period. The seller incurs a cost of c as per unit purchasing cost and earns a selling price of p for each unit sold. During the selling period, if demand is more than the order quantity, then he will incur cost in terms of lost sales $C_u$ = p-c and if demand is less than the order quantity, he will have more in stock and will incur the cost of overstocking $C_o$ = c-s (assume the salvage value is zero). If the demand distribution function is F and the density function is f, the realized profit of the seller when he orders quantity q and demand is D is:

$$\pi \ (q, D) = p \min(q, D) - cq$$

And the seller's expected profit is:

$$\underline{E}[\pi(q,D)] = (1 - F(q)\pi(q,q) + \int_0^q f(x)\pi(q,x)dx \quad \text{(A)}$$

## 2.3 Metrics for Analysis

Two measures will be used to measure the performance of the subjects: Average order quantity and average expected profit. The average order quantity Qi for each subject(i) will calculated over n rounds of ordering:

$$Qi = 1/n \sum_{j=1}^n qj$$

where q(j) is the order quantity of round j.

The average expected profit will be calculated for the subject over n round of ordering:

$$Epi = 1/n \sum_{j=1}^n E[\pi(qj)]$$

Where E[π(qj)] is the expected profit given by equation(A).

## 3. RESEARCH HYPOTHESIS

### 3.1 Construal Level and Order Quantity and Expected Profit

In goal-directed actions, the level of mental construal allows people to construe goals at different levels and individuals take action based on whether the goal is construed at a superordinate level or subordinate level (Muzumdar, 2014c). For example, conducting a study will be construed as the advancement of science by abstract thinkers and just entering the data by concrete thinkers (Muzumdar, 2012c). This implies that in the newsvendor setting, abstract thinkers will construe a superordinate goal that is to achieve the optimal quantity and higher profit both in high and low-margin settings. On the other hand, concrete thinkers will construe a subordinate goal that is to meet demand with consideration of avoiding too much or too little on high profit and low margin setting respectively.

Abstract thinkers will focus on the preferences that will maximize the main aim of the decision whereas concrete thinkers will lean toward the ease of process of decision (Mark et al., 2021). Kurian & Muzumdar (2018) argued that individuals with abstract thinking will take a holistic view of the situation and will be able to extract valuable information from the environment that will lead to enhanced performance as compared to concrete thinkers. The results of tasks involving insight are better if the task is construed at an abstract level as solving of the problem is eased out by interpreting it at an abstract level (Muzumdar, 2022b). We would argue that news vendor problems need insight to arrive at optimal solutions, therefore subjects with abstract thinking performance will be better than that of concrete thinkers (Muley et al., 2023). In addition, individuals with a high level of construal refer to why-oriented questions, and individuals with low-level construal refer to how-orientated questions during decision-making (Mark et al., 2021). In the newsvendor game, abstract thinkers will refer to questions such as why it is important to consider demand distribution, price, and cost of the unit and what implication it will have on the supply chain performance. On the other hand, concrete thinkers will consider questions such as how much to order without considering the overall implication of their decision on supply chain performance.

Based on the above arguments, the following hypotheses are proposed:

**Hypothesis 1: Individuals with abstract thinking will have order quantities closer to optimal quantity and higher expected profit than individuals with concrete thinking both in high-margin and low-margin settings.**

**Hypothesis 2: The order variance will be lower for individuals with abstract thinking processes than for individuals with concrete thinking both in high-margin and low-margin settings.**

## 3.2 Demand Chasing

High-level construal's assist in thinking of events as been carrying out under situations that are dissimilar from those currently experienced (Muzumdar, 2015c). They further argued that it allows individuals with high levels of construal to broaden their perspective. This allows individuals having abstract thinking to surpass "here and now" whereas individuals with low level construal will focus on the "here and now" aspect of the event. Authors argued that the level of construal affects creativity and it has an impact on decision making (Muzumdar, 2022a). In addition, they suggest that people having high-level construal are more creative and they use more nontraditional ways to approach the problem . People with high level of construal try to avoid base rate fallacy and other biases that arises due to relying on heuristics.

In difficult situations, people tend to use shortcuts and heuristics (Muzumdar, 2021b). Kurian & Muzumdar (2020) argued that people using abstract thinking do not depend on contextual information that allows them to diminish the judgment bias and abstract thinking may act as a debiasing strategy. In the newsvendor setting, the literature suggests that individuals involved in demand-chasing behavior (see (Muzumdar, 2013 & Muzumdar, 2011) which implies individuals' decisions are based on the realized demand from the previous period. The above arguments imply that in a news vendor setting, people with abstract thinking will be less prone to demand chasing and individuals with concrete thinking will be more prone to demand chasing, and based on the above argument following hypothesis is proposed:

**Hypothesis 3: In a news vendor setting, Individuals with concrete thinking will demonstrate more demand-chasing behavior based on the previous period than individuals with high abstract thinking.**

# 4. MATERIALS AND METHODS

## 4.1 Laboratory Setting

In this study, both high-margin conditions and low-margin conditions will be considered. The parameter for this study will be adopted from the study (Muzumdar, 2021a). In low margin setting the selling price ( p) \$12 per unit and the cost per unit (c) is \$9. The demand distribution (Uniform) in the low margin will be D ~(0,100) with an optimal quantity of 25. In high margin setting the price per unit p is \$12 and the cost per unit (c) is \$3. The demand distribution in the high margin setting will be D~(0,100) and the optimal quantity in the high margin setting will be 75.

## 4.2 Participants

100 participants will be selected from undergraduate class of business school from a major university in the west south region of the United States. Before the random assignment to different treatment groups, a cognitive reflection test (CRT) will be administered to measure the cognitive reflection of each individual (details about CRT are discussed in the next section). The participants will be randomly assigned to high and low-margin conditions and then participants will be randomly assigned to abstract condition and concrete conditions. The participants will gather in the behavior lab and all participants will be explained about the newsvendor problem but they will not be informed about the optimal order quantity and expected profit maximization. Each participant will play the game for a hundred rounds of newsvendor game and demand will be drawn randomly and independently from the demand distribution. The game will be played on the computer and game will be designed by the author. All participants will be paid in cash. The base rate to participate in the game will be \$10 for each participant and they can earn subsequent money based on their performance in the game.

### 4.3 Cognitive Reflection Test

Mortiz et al. (2013) argued that the cognitive reflection of individuals may explain some of the variation in newsvendor settings. They measured the cognitive reflection using the cognitive reflection test (Zare et al., 2023) and found some support. They were able to show variation in the high margin setting but found no evidence in the low margin setting in the newsvendor problem. Researchers have suggested that CRT is one of the potent predictors of performance on heuristics and bias tasks (Muzumdar & Kurian, 2019a). In this study, the main aim is to examine the impact of the level of mental construal on performance in a news vendor setting and cognitive reflection of individual may confound the results. Although these two concepts measure totally different traits, some of the variance may overlap. In order to study the true impact of construal level, the cognitive reflection of individuals will measure ex-ante to control any shared variance between construal level and cognitive reflection. They developed a cognitive reflection test (CRT) to measure the cognitive reflection of individuals and in this paper, the same CRT test will be administered to measure the cognitive reflection of each individual (Kurian & Muzumdar, 2017b). The CRT test is composed of three quantitative items (see Table 2) and the CRT score of each individual will be the sum of correct answers on the cognitive reflection test (CRT).

### 4.4 Priming of Subjects

In this study, subjects will be primed to think abstractly or concretely. The concepts of priming are well established in social psychology experiments. Goals can be primed passively and non-consciously and results are the same to that of if the goals are followed consciously and mental and information processing goals can be triggered non-consciously by priming the subjects (Muzumdar, 2014a). They did five experiments and found that behavioral goals can be triggered unconsciously and these nonconscious goals works in same way as for consciously selected goal (Muzumdar, 2015a).

They used priming to think individually at the abstract level and concrete level. In this study, we will adopt the exercise based on design (Kurian & Muzumdar, 2017a). Participants will be randomly assigned to either abstract or concrete conditions. In abstract thinking, subjects will direct to think about the why aspect of the activity, and in concrete thinking, they will be directed to think how aspect of the activity. The activity will be common for all participants so that content domain and decision status are constant.

Participants in the abstract condition will be asked to consider why they would engage in a particular activity. At the start of the exercise, all the participants in abstract condition will read a passage that describe why a person want to do an activity. The passage appears below for everything we do, there is always a reason why we do it. Moreover, we often can trace the causes of our behavior back to broad life goals that we have. For example, you currently are participating in an operation behavior experiment. Why are you doing this? Perhaps to satisfy a course requirement. Why are you satisfying the course requirement? Perhaps to pass an operation management course. Why pass the course? Perhaps because you want to earn a college degree. Why earn a college degree? Maybe because you want to find a good job, or because you want to educate yourself. Perhaps you wish to educate yourself or find a good job because you feel that doing so can bring you happiness in life. Research suggests that engaging in thought exercises like that above, in which one thinks about how one's actions relate to one's ultimate life goal. In this experiment, we are testing such a technique. This thought exercise is intended to focus your attention on why you do the things you do.

**Table 2. Cognitive reflection test instrument**

| Q1. A bat and a ball cost $1.10 in total. The bat costs $1 more than the ball. How much does the ball cost? ____ cents |
| --- |
| Q2. If it takes 5 machines 5 min to make 5 widgets, how long would it take 100 machines to make 100 widgets? ____ minutes |
| Q3. In a lake, there is a patch of lily pads. Every day, the patch doubles in size. If it takes 48 days for the patch to cover the entire lake, how long would it take for the patch to cover half the lake? ____ days |

For this exercise, please consider the following activity, you are an inventory manager at XYZ company and you are responsible for managing the inventory of a product whose demand is uncertain and optimal quantity is crucial for the financial health of your organization (Reavis et al., 2024). So, your goal is, "to manage the order quantity of the product,". The subjects assigned to the abstract condition will asked the following question, "Why do they need to manage the order quantity,". The participants will be presented with the diagram as shown in Fig. 1 that is labeled why, "to manage the order quantity of the product" on the bottom box of the diagram, and then subjects will be asked to fill in the answer in the box immediate above the bottom box of why they need to manage the order quantity of the product. On completing that answer, participants will be again prompted to ask themselves why they need to do that. For example, the first response of participants can be, "to better serve the customer and improve the financial position of the organization". The diagram will prompt them to ask themselves "Why do they need to serve the customer better". In this way, the subjects have to provide four responses. Participants in concrete conditions will be asked to consider how they would engage in a particular activity. At the start of the exercise, all the participants in concrete conditions will read a passage that describes how a person wants to do an activity. The passage appears below:

For everything we do, there always is a process of how we do it. Moreover, we often can follow our broad life goals down to our very specific behaviors. For example, like most people, you probably hope to do well in life. How can you do this? Perhaps finding a good job, or being educated, can help. How can you do these things? Perhaps by earning a college degree. How do you earn a college degree? By satisfying course requirements. How do you satisfy course requirements? In some cases, such as today, you participate in an operation behavior experiment. Research suggests that engaging in a thought exercise like that above, in which one thinks about how one's ultimate life goals can be expressed through specific actions. In this experiment, we are testing such a technique.

This thought exercise is intended to focus your attention on how you do the things you do. For this exercise, please consider the following activity, you are an inventory manager at XYZ company and you are responsible for managing the inventory of a product whose demand is uncertain and optimal quantity is crucial for the financial health of your organization. So your goal is, "to manage the order quantity of the product,". The subject who is assigned to low-level construal will be asked the question, "To manage the order quantity of the product". The subject will be presented with a vertical diagram box similar to the abstract condition, but the boxes will start from the top of the page (see Fig 1).

The top box will fill with a line, "to manage the order quantity of the product". The participants will be asked to write a response on how they will manage the order quantity of the product and write their response in the box just below the top box. After writing their first response, the participants will be prompted to ask how they would engage in their first response. For example, the first response can be," I will look at the current stock level "and then the second question based on this response will be how you will look at the current stock level. In this setting, subjects will provide four responses.

## 4.5 Manipulation Check

To ascertain that the priming of subjects worked as anticipated, a manipulation check will conducted and a method will be used to check the manipulation. The responses (why or how) of the participants will be evaluated by two experts who are not aware of the condition. The responses will be checked against the criterion Y by X, where X is the response of the subject to prompt Y. For example, if the participant's responses are subordinate means to original statement, "to manage the order quantity of the product," the judges will code the response of participant with a score of -1. If the participant's responses are superordinate ends to the original statement, "to manage the order quantity of the product," the judges will code the response of the participant with a score of +1. If the response of the participant does fit any of these two criteria, the judges will code the response as 0. The rating of the subject will be calculated for all four responses and then it will be summed to create an index for the level of construal and the range of score will be between -4 and +4. In addition, the inter-rater reliability will also be checked to see the agreement of both judges.

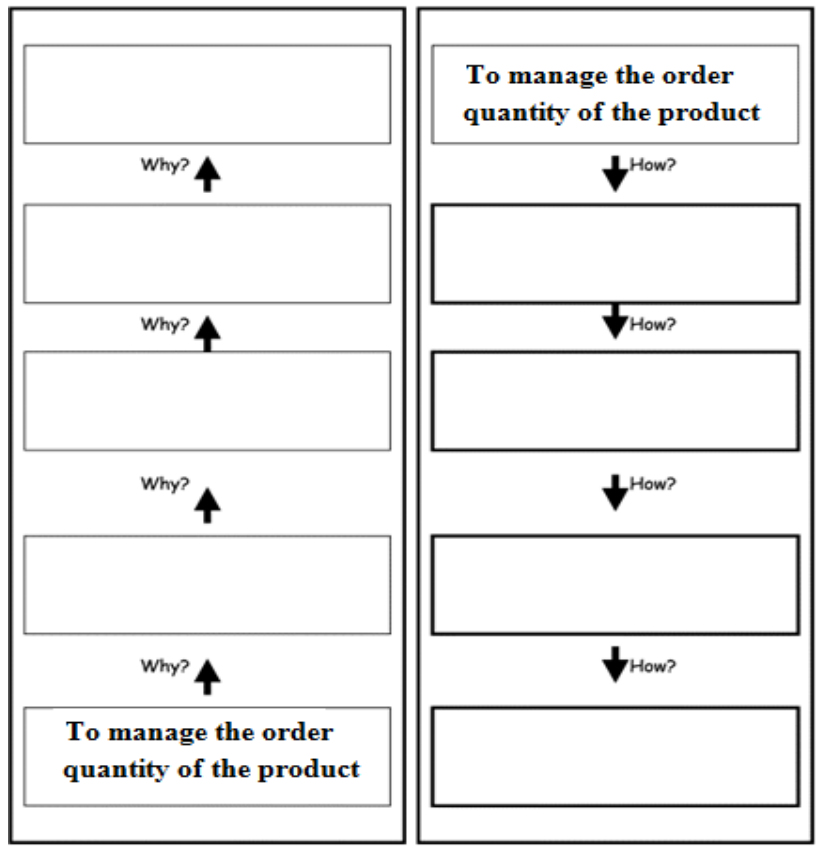

**Fig. 1. The vertical diagram box**

## 5. RESULTS AND DISCUSSION

The discussion assumes that all hypotheses are true. The result of this study implies that how the decision maker construes a problem either in abstract form or concrete form influences the optimal quantity under uncertainty. This study contributes to previous behavior research in news vendor settings and extends the previous work by extracting some variation in decision-making.

The contributions of this study are twofold, First, this is the first study to use construal level theory (CLT) to study individual variation in a news vendor setting. This study highlights that subject's use different cognitive mechanisms to view the problem and that results in deviation in decision making. Previous studies on news vendor problems have shown that people deviate from normative solutions, but they were not able to establish the cause of it except few (Limaye et al., 2023). The results of this study highlight the underlying mental process which results in non-optimal solutions by the decision-makers. In this study, it is shown that people who use abstract thinking are more capable of making better decisions, close to normative solutions as compared to those who use concrete thinking (Basyal et al., 2020). The managerial implication of this study is that human resource people should devise a test to measure the level of construal level at the time of hiring people in the supply chain or at the time of promoting the person to a position that requires decision-making and that is critical for the organization. Moreover, organizations can develop specialized training programs for those individuals or departments in which they are required to think abstractly to achieve the desired result.

## 6. CONCLUSION

In this study, the relationship between the level of construal and its impact on performance in news vendor setting is examined. This study highlights some of the variations in decision making that can be attributed to how decision-makers construe a problem in the mind and how it impacts the performance in a news vendor setting.

## DISCLAIMER (ARTIFICIAL INTELLIGENCE)

Author(s) hereby declare(s) that NO generative AI technologies such as Large Language Models (ChatGPT, COPILOT, etc.) and text-to-image generators have been used during the writing or editing of this manuscript.

## COMPETING INTERESTS DISCLAIMER

Authors have declared that they have no known competing financial interests or non-financial interests or personal relationships that could have appeared to influence the work reported in this paper.

---

*Peer-review history:*
*The peer review history for this paper can be accessed here:*
*https://www.sdiarticle5.com/review-history/129556*

---